\pdfoutput=1
\documentclass[12pt,a4paper]{article}
\usepackage{ifthen} % for conditional statements
\newboolean{pdflatex}
\setboolean{pdflatex}{true} % False for eps figures 

\newboolean{articletitles}
\setboolean{articletitles}{true} % False removes titles in references

\newboolean{uprightparticles}
\setboolean{uprightparticles}{false} %True for upright particle symbols

\newboolean{inbibliography}
\setboolean{inbibliography}{false} %True once you enter the bibliography

\usepackage{microtype}
\usepackage{lineno}  % for line numbering during review
\usepackage{xspace} % To avoid problems with missing or double spaces after
\usepackage{caption} %these three command get the figure and table captions automatically small

\usepackage{graphicx}  % to include figures (can also use other packages)
\usepackage{color}
\usepackage{colortbl}
\graphicspath{{./figs/}} % Make Latex search fig subdir for figures

\usepackage{amsmath} % Adds a large collection of math symbols
\usepackage{amssymb}
\usepackage{amsfonts}
\usepackage{upgreek} % Adds in support for greek letters in roman typeset

\newcommand*\patchAmsMathEnvironmentForLineno[1]{%
\expandafter\let\csname old#1\expandafter\endcsname\csname #1\endcsname
\expandafter\let\csname oldend#1\expandafter\endcsname\csname
end#1\endcsname
 \renewenvironment{#1}%
   {\linenomath\csname old#1\endcsname}%
   {\csname oldend#1\endcsname\endlinenomath}%
}
\newcommand*\patchBothAmsMathEnvironmentsForLineno[1]{%
  \patchAmsMathEnvironmentForLineno{#1}%
  \patchAmsMathEnvironmentForLineno{#1*}%
}
\AtBeginDocument{%
\patchBothAmsMathEnvironmentsForLineno{equation}%
\patchBothAmsMathEnvironmentsForLineno{align}%
\patchBothAmsMathEnvironmentsForLineno{flalign}%
\patchBothAmsMathEnvironmentsForLineno{alignat}%
\patchBothAmsMathEnvironmentsForLineno{gather}%
\patchBothAmsMathEnvironmentsForLineno{multline}%
\patchBothAmsMathEnvironmentsForLineno{eqnarray}%
}

\usepackage{hyperref}    % Hyperlinks in references
\usepackage[all]{hypcap} % Internal hyperlinks to floats.

\def\lhcb {\mbox{LHCb}\xspace}

\def\MagUp {\mbox{\em Mag\kern -0.05em Up}\xspace}

\ifthenelse{\boolean{uprightparticles}}%
{

 \def\Pmu         {\ensuremath{\upmu}\xspace}

 \def\Ppi         {\ensuremath{\uppi}\xspace}

 \def\PDelta      {\ensuremath{\Delta}\xspace}                 
 \def\PXi      {\ensuremath{\Xi}\xspace}                 
 \def\PLambda      {\ensuremath{\Lambda}\xspace}                 
 \def\PSigma      {\ensuremath{\Sigma}\xspace}                 
 \def\POmega      {\ensuremath{\Omega}\xspace}                 
 \def\PUpsilon      {\ensuremath{\Upsilon}\xspace}                 
 
 \def\PB      {\ensuremath{\mathrm{B}}\xspace}                 
                  
 \def\PD      {\ensuremath{\mathrm{D}}\xspace}

 \def\PK      {\ensuremath{\mathrm{K}}\xspace}

 \def\Pb      {\ensuremath{\mathrm{b}}\xspace}                 
 \def\Pc      {\ensuremath{\mathrm{c}}\xspace}

 \def\Pi      {\ensuremath{\mathrm{i}}\xspace}

}
{

 \def\Pmu         {\ensuremath{\mu}\xspace}

 \def\Ppi         {\ensuremath{\pi}\xspace}

 \mathchardef\PDelta="7101
 \mathchardef\PXi="7104
 \mathchardef\PLambda="7103
 \mathchardef\PSigma="7106
 \mathchardef\POmega="710A
 \mathchardef\PUpsilon="7107
                  
 \def\PB      {\ensuremath{B}\xspace}                 
                  
 \def\PD      {\ensuremath{D}\xspace}

 \def\PK      {\ensuremath{K}\xspace}

 \def\Pb      {\ensuremath{b}\xspace}                 
 \def\Pc      {\ensuremath{c}\xspace}

 \def\Pi      {\ensuremath{i}\xspace}

}

\makeatletter
\ifcase \@ptsize \relax% 10pt
  \newcommand{\miniscule}{\@setfontsize\miniscule{4}{5}}% \tiny: 5/6
\or% 11pt
  \newcommand{\miniscule}{\@setfontsize\miniscule{5}{6}}% \tiny: 6/7
\or% 12pt
  \newcommand{\miniscule}{\@setfontsize\miniscule{5}{6}}% \tiny: 6/7
\fi
\makeatother

\DeclareRobustCommand{\optbar}[1]{\shortstack{{\miniscule (\rule[.5ex]{1.25em}{.18mm})}
  \\ [-.7ex] $#1$}}

\def\mup        {{\ensuremath{\Pmu^+}}\xspace}
\def\mun        {{\ensuremath{\Pmu^-}}\xspace} % muon negative (\mum is taken)

\def\cquark    {{\ensuremath{\Pc}}\xspace}

\def\bquark    {{\ensuremath{\Pb}}\xspace}

\def\pion   {{\ensuremath{\Ppi}}\xspace}
\def\piz    {{\ensuremath{\pion^0}}\xspace}

\def\kaon    {{\ensuremath{\PK}}\xspace}
  \def\Kbar    {{\kern 0.2em\overline{\kern -0.2em \PK}{}}\xspace}

\def\KorKbar    {\kern 0.18em\optbar{\kern -0.18em K}{}\xspace}

\def\Kpm     {{\ensuremath{\kaon^\pm}}\xspace}

  \def\Dbar    {{\kern 0.2em\overline{\kern -0.2em \PD}{}}\xspace}

\def\DorDbar    {\kern 0.18em\optbar{\kern -0.18em D}{}\xspace}

\def\B       {{\ensuremath{\PB}}\xspace}
\def\Bbar    {{\ensuremath{\kern 0.18em\overline{\kern -0.18em \PB}{}}}\xspace}

\def\BorBbar    {\kern 0.18em\optbar{\kern -0.18em B}{}\xspace}

\def\Bpm     {{\ensuremath{\B^\pm}}\xspace}

  \def\Y#1S{\ensuremath{\PUpsilon{(#1S)}}\xspace}% no space before {...}!

\def\Lbar        {{\ensuremath{\kern 0.1em\overline{\kern -0.1em\PLambda}}}\xspace}
\def\LorLbar    {\kern 0.18em\optbar{\kern -0.18em \PLambda}{}\xspace}

\def\to                 {\ensuremath{\rightarrow}\xspace}

\def\AT#1     {\ensuremath{A_{\mathrm{T}}^{#1}}\xspace}           % 2

\def\C#1      {\ensuremath{\mathcal{C}_{#1}}\xspace}                       % 9
\def\Cp#1     {\ensuremath{\mathcal{C}_{#1}^{'}}\xspace}                    % 7
\def\Ceff#1   {\ensuremath{\mathcal{C}_{#1}^{\mathrm{(eff)}}}\xspace}        % 9  
\def\Cpeff#1  {\ensuremath{\mathcal{C}_{#1}^{'\mathrm{(eff)}}}\xspace}       % 7
\def\Ope#1    {\ensuremath{\mathcal{O}_{#1}}\xspace}                       % 2
\def\Opep#1   {\ensuremath{\mathcal{O}_{#1}^{'}}\xspace}                    % 7

\newcommand{\tev}{\ifthenelse{\boolean{inbibliography}}{\ensuremath{~T\kern -0.05em eV}\xspace}{\ensuremath{\mathrm{\,Te\kern -0.1em V}}}\xspace}
\newcommand{\gev}{\ensuremath{\mathrm{\,Ge\kern -0.1em V}}\xspace}
\newcommand{\mev}{\ensuremath{\mathrm{\,Me\kern -0.1em V}}\xspace}
\newcommand{\kev}{\ensuremath{\mathrm{\,ke\kern -0.1em V}}\xspace}
\newcommand{\ev}{\ensuremath{\mathrm{\,e\kern -0.1em V}}\xspace}
\newcommand{\gevc}{\ensuremath{{\mathrm{\,Ge\kern -0.1em V\!/}c}}\xspace}
\newcommand{\mevc}{\ensuremath{{\mathrm{\,Me\kern -0.1em V\!/}c}}\xspace}
\newcommand{\gevcc}{\ensuremath{{\mathrm{\,Ge\kern -0.1em V\!/}c^2}}\xspace}
\newcommand{\gevgevcccc}{\ensuremath{{\mathrm{\,Ge\kern -0.1em V^2\!/}c^4}}\xspace}
\newcommand{\mevcc}{\ensuremath{{\mathrm{\,Me\kern -0.1em V\!/}c^2}}\xspace}

\def\invpb {\ensuremath{\mbox{\,pb}^{-1}}\xspace}

\def\invfb   {\ensuremath{\mbox{\,fb}^{-1}}\xspace}

\def\gsim{{~\raise.15em\hbox{$>$}\kern-.85em
          \lower.35em\hbox{$\sim$}~}\xspace}
\def\lsim{{~\raise.15em\hbox{$<$}\kern-.85em
          \lower.35em\hbox{$\sim$}~}\xspace}

\def\sPlot{\mbox{\em sPlot}\xspace}
\def\ptot       {\mbox{$p$}\xspace}
\def\pt         {\mbox{$p_{\rm T}$}\xspace}
\def\ptsq     {\mbox{$p^2_{\rm T}$}\xspace}

\def\evtgen     {\mbox{\textsc{EvtGen}}\xspace}

\def\geant      {\mbox{\textsc{Geant4}}\xspace}

\def\photos     {\mbox{\textsc{Photos}}\xspace}

\def\tell1  {TELL1\xspace}
\def\ukl1   {UKL1\xspace}

\usepackage{cite} % Allows for ranges in citations
\usepackage{mciteplus}

\usepackage{subfigure}
\begin{document}

%%%%%%%%%%%%%%%%%%%%%%%%%
%%%%% Title     %%%%%%%%%
%%%%%%%%%%%%%%%%%%%%%%%%%
\renewcommand{\thefootnote}{\fnsymbol{footnote}}
\setcounter{footnote}{1}

% %%%%%%% CHOOSE TITLE PAGE--------
%\onecolumn
% \input{title-LHCb-ANA}
%\input{title-LHCb-CONF}
% $Id: title-LHCb-PAPER.tex 65671 2015-01-09 14:59:08Z tgershon $
% ===============================================================================
% Purpose: LHCb-PAPER journal paper title page template
% Author: 
% Created on: 2010-09-25
% ===============================================================================

%%%%%%%%%%%%%%%%%%%%%%%%%
%%%%%  TITLE PAGE  %%%%%%
%%%%%%%%%%%%%%%%%%%%%%%%%
\begin{titlepage}
\pagenumbering{roman}

% Header ---------------------------------------------------
\vspace*{-1.5cm}
\centerline{\large EUROPEAN ORGANIZATION FOR NUCLEAR RESEARCH (CERN)}
\vspace*{1.5cm}
\hspace*{-0.5cm}
\begin{tabular*}{\linewidth}{lc@{\extracolsep{\fill}}r}
\ifthenelse{\boolean{pdflatex}}% Logo format choice
{\vspace*{-2.7cm}\mbox{\!\!\!\includegraphics[width=.14\textwidth]{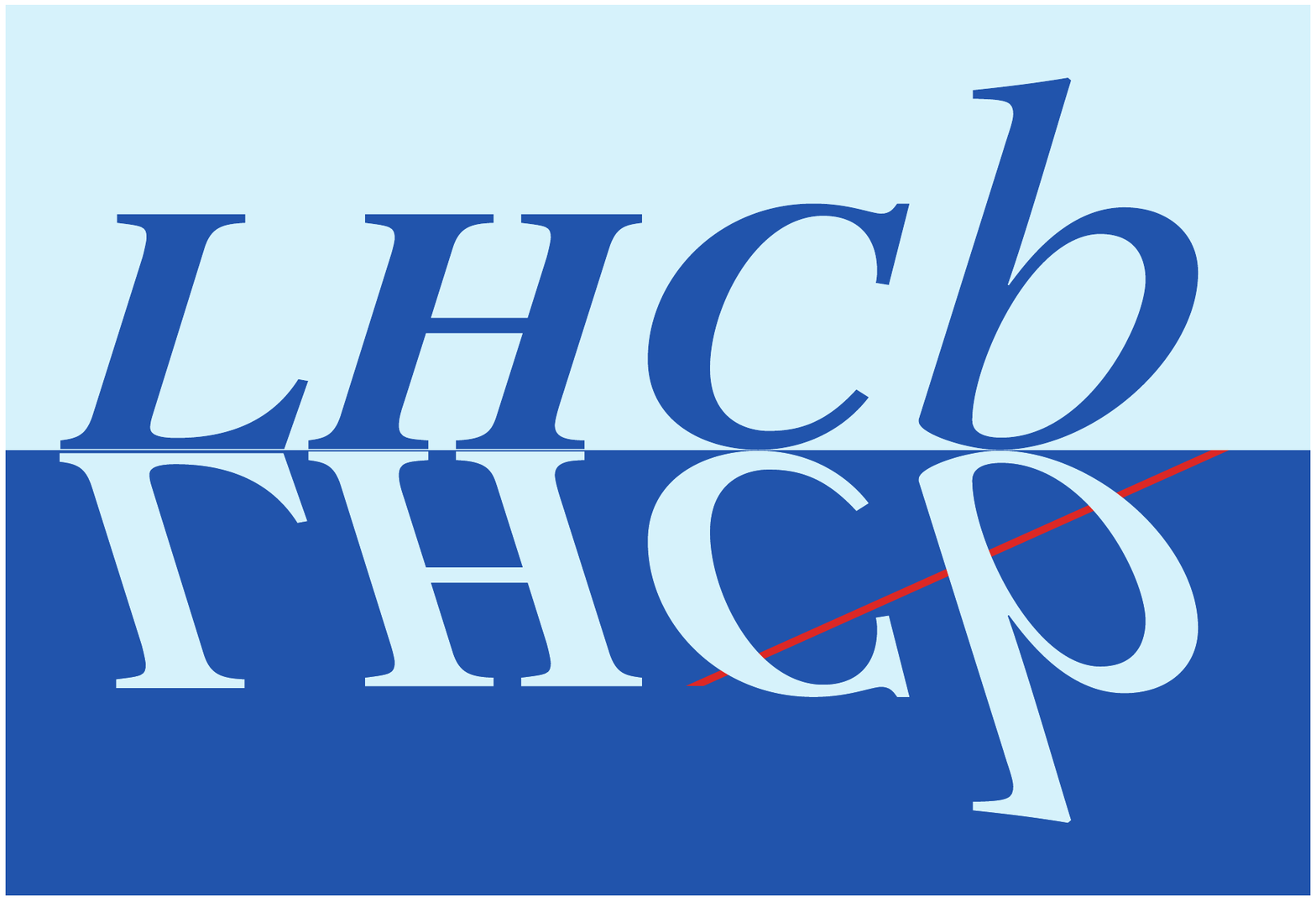}} & &}%
{\vspace*{-1.2cm}\mbox{\!\!\!\includegraphics[width=.12\textwidth]{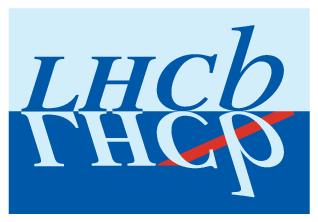}} & &}%
\\
 & & CERN-PH-EP-2015-123 \\  % ID 
 & & LHCb-PAPER-2015-011 \\  % ID 
 & & September 21, 2015 \\ % Date - Can also hardwire e.g.: 23 March 2010
 & & \\
% not in paper \hline
\end{tabular*}

\vspace*{4.0cm}

% Title --------------------------------------------------
{\bf\boldmath\huge
\begin{center}
Measurement of the exclusive $\Upsilon$ production cross-section in $pp$ collisions at $\sqrt{s}=7\tev$ and 8\tev
\end{center}
}

\vspace*{0.5cm}

% Authors -------------------------------------------------
\begin{center}
%In the footnote, replace 'paper' by 'letter' in case of submission to PRL or PLB 
The LHCb collaboration\footnote{Authors are listed at the end of this paper.}
\end{center}

%\vspace{\fill}

% Abstract -----------------------------------------------
\begin{abstract}
  \noindent
A study is presented of central exclusive production of $\Upsilon(nS)$ states, where the $\Upsilon(nS)$ resonances decay to the $\mup\mun$ final state, using $pp$ collision data recorded by the LHCb experiment. The cross-section is measured in the rapidity range $2<y(\Upsilon)<4.5$ where the muons are reconstructed in the pseudorapidity range $2<\eta(\mu^\pm)<4.5$. The data sample corresponds to an integrated luminosity of 2.9\invfb and was collected at centre-of-mass energies of $7\tev$ and $8\tev$. The measured $\Upsilon(1S)$ and $\Upsilon(2S)$  production cross-sections are
\begin{eqnarray}
\sigma(pp \to p\Upsilon(1S)p) &=& 9.0 \pm 2.1 \pm 1.7\textrm{ pb and}\nonumber\\
\sigma(pp \to p\Upsilon(2S)p) &=& 1.3 \pm 0.8 \pm 0.3\textrm{ pb},\nonumber
\end{eqnarray}
where the first uncertainties are statistical and the second are systematic. 
The $\Upsilon(1S)$ cross-section is also measured as a function of rapidity and is found to be in good agreement with Standard Model predictions. 
An upper limit is set at  3.4\,pb at the 95\% confidence level for the exclusive $\Upsilon(3S)$ production cross-section, including possible contamination from $\chi_b(3P)\to\Upsilon(3S)\gamma$ decays.
\end{abstract}

%\vspace*{2.0cm}

\begin{center}
  Published in JHEP
\end{center}

\vspace{\fill}

{\footnotesize 
\centerline{\copyright~CERN on behalf of the \lhcb collaboration, licence \href{http://creativecommons.org/licenses/by/4.0/}{CC-BY-4.0}.}}
\vspace*{2mm}

\end{titlepage}

%%%%%%%%%%%%%%%%%%%%%%%%%%%%%%%%
%%%%%  EOD OF TITLE PAGE  %%%%%%
%%%%%%%%%%%%%%%%%%%%%%%%%%%%%%%%

%  empty page follows the title page ----
\newpage
\setcounter{page}{2}
\mbox{~}
%\newpage
%
%% Author List ----------------------------
%%  You need to get a new author list!
%\input{LHCb_authorlist.tex}
%
%The author list for journal publications is provided by the Membership Committee shortly after 'approval to go to paper' has been given.
%%It will be made available on the page
%%\verb!http://www.physik.uzh.ch/~strauman/forMemCo/LHCb-PAPER-XXXX-XXX/! .
%It will be sent to you by email shortly after a paper number has beens assigned.
%The author list should be included already at first circulation, 
%to allow new members of the collaboration to verify whether they have been included correctly.
%Occasionally a misspelled name is corrected or associated institutions become full members.
%In that case, a new author list will be sent to you.
%In case line numbering doesn't work well after including the authorlist, try moving the \verb!\bigskip! after the last author to a separate line.
%
%
%The authorship for Conference Reports should be ``The LHCb
%  collaboration'', with a footnote giving the name(s) of the contact
%  author(s), but without the full list of collaboration names.

\cleardoublepage

%\twocolumn
% %%%%%%%%%%%%% ---------

\renewcommand{\thefootnote}{\arabic{footnote}}
\setcounter{footnote}{0}

%%%%%%%%%%%%%%%%%%%%%%%%%%%%%%%%
%%%%%  Table of Content   %%%%%%
%%%%%%%%%%%%%%%%%%%%%%%%%%%%%%%%
%%%% Uncomment next 2 lines if desired
%\tableofcontents
%\cleardoublepage

%%%%%%%%%%%%%%%%%%%%%%%%%
%%%%% Main text %%%%%%%%%
%%%%%%%%%%%%%%%%%%%%%%%%%

\pagestyle{plain} % restore page numbers for the main text
\setcounter{page}{1}
\pagenumbering{arabic}

%% Uncomment during review phase. 
%% Comment before a final submission.
%\linenumbers

% You can include short sections directly in the main tex file.
% However, for larger papers it is desirable to split the text into
% several semiautonomous files, which can be revised independently.
% This is especially useful when developing a document in
% collaboration with several people, since then different parts can be
% edited independently.  This type of file organization is shown here.
% 

% $Id: introduction.tex 65669 2015-01-09 14:55:20Z tgershon $

\section{Introduction}
\label{sec:Introduction}

\begin{figure}[b]
\centering
\includegraphics[width=.43\textwidth]{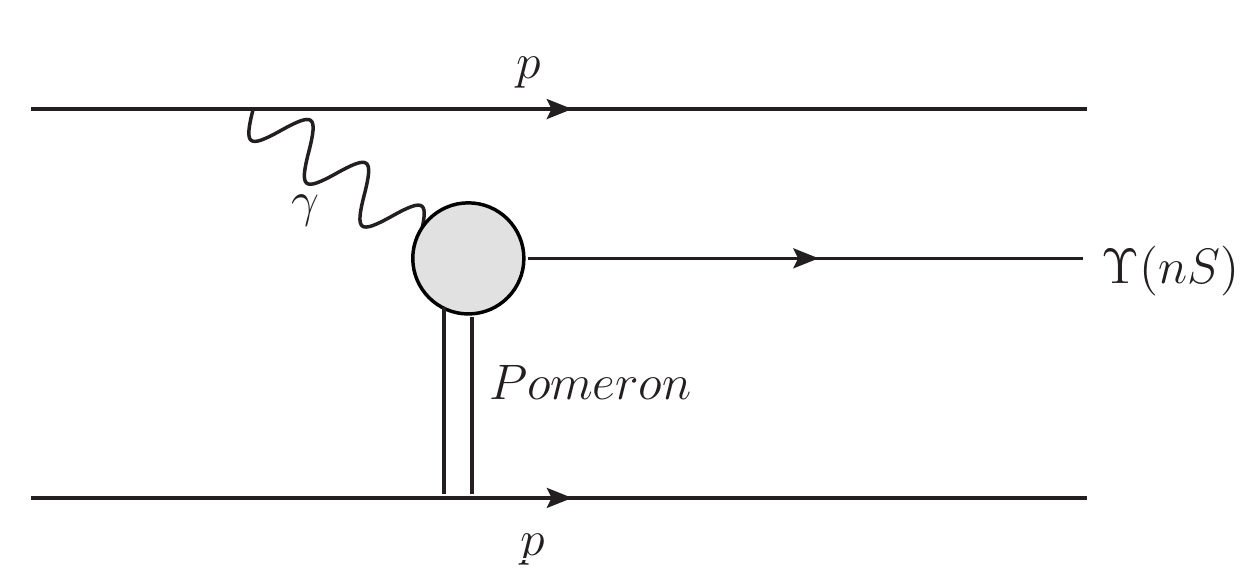}
\caption{Leading Feynman diagram for photoproduction of $\Upsilon(nS)$ states, where the photon-pomeron interaction is indicated by the shaded grey circle.\label{im:FeynDiag}}
\end{figure}

Central exclusive production (CEP) of $\Upsilon(nS)$ $(n=1,2,3)$ resonances in $pp$ collisions is thought to occur by photoproduction through the exchange of a photon and a pomeron (a colour-singlet system) between two protons, as illustrated in Fig.~\ref{im:FeynDiag}. Since the protons do not dissociate, typically only a small component of momentum transverse to the beam direction (\pt) is exchanged in the interaction. The photoproduction of $\Upsilon$ resonances at LHCb can be computed using perturbative quantum chromodynamics (QCD), given the high photon-proton centre-of-mass energy, $W$, and the cross-section depends on the square of the gluon parton-density function, $g(x)$, where Bjorken-$x$ is the fraction of the proton's momentum carried by the gluon \cite{Jones:2013pga}. Measurements of the production cross-sections for the $\Upsilon(nS)$ resonances in the forward region covered by the LHCb detector are sensitive to $g(x)$ in the region of small $x$ down to approximately $1.5\times 10^{-5}$, where the knowledge of $g(x)$ is limited. Furthermore, predictions for the $\Upsilon(nS)$ cross-sections at leading order (LO) and next-to-leading order (NLO) in the strong-interaction coupling differ greatly for the values of $W$ probed in $\Upsilon(nS)$ resonance production, and there are significant variations depending on the models used to describe the $\Upsilon$ wave function and the $t$-channel exchange \cite{Jones:2013pga,GoncalvesUpsilon2014,Dosch:2015jua}.

Quarkonia photoproduction has been studied in exclusive production at HERA \cite{ZEUS1998,HERAJpsiAndUpsilonExclProduction2000,ZEUSJpsi2002,ZEUS2009,ZEUSUpsilontDep,HERAJpsi2013}, the Tevatron \cite{CDFExclusives} and the LHC \cite{ALICEPbPb2012,RonanUpdatedJpsi,ALICEpPb2014}. At LHCb, exclusive production is associated with the absence of significant detector activity apart from that associated with the exclusive candidate. The background from proton dissociation occurring outside the detector acceptance is characterised as having a value of $\Upsilon$ candidate $p_T$ which is larger than that for exclusive production.

In this article, the exclusive production cross-section of $\Upsilon(nS)$ resonances is measured in the $\mup\mun$ final state where both muons lie in the pseudorapidity ($\eta$) range $2<\eta(\mu^\pm)<4.5$ and the $\Upsilon(nS)$ candidate is reconstructed in the rapidity ($y$) range $2< y(\Upsilon(nS))<4.5$. The $pp$ data correspond to an integrated luminosity of 0.9\invfb at a $pp$ centre-of-mass energy of $\sqrt{s}=7\tev$ and 2.0\invfb at $\sqrt{s}=8\tev$. Given the limited statistical precision, the data sets are combined to measure the production cross-sections. The LHCb detector and the simulated event samples are outlined in Sect.~\ref{sec:Detector}. In Sect.~\ref{sec:selectionAndEfficiency} selection criteria are discussed, which exploit the absence of detector activity other than that associated with the $\Upsilon(nS)$ candidate. The signal efficiency and the various sources of background are also described. In Sect.~\ref{sec:Fitting} two fits are described, which allow the determination of the exclusive signal yield: by fitting the $\Upsilon(nS)$ invariant mass spectrum in order to separate $\Upsilon$ resonances from dimuon continuum background; and by fitting the $\Upsilon(nS)$ candidate $\ptsq$ distribution to distinguish exclusively produced $\Upsilon$ resonances from those originating in hard interactions. Systematic uncertainties are summarised in Sect.~\ref{sec:Systematics}, and the measurements of the cross-sections are discussed in Sect.~\ref{sec:CrossSection}. Finally, for the $\Upsilon(1S)$ the differential cross-section, as a function of $\Upsilon(1S)$ candidate rapidity, is presented.

\section{Detector and simulation}
\label{sec:Detector}
The \lhcb detector~\cite{Alves:2008zz,LHCb-DP-2014-002} is a single-arm forward
spectrometer,
designed for the study of particles containing \bquark or \cquark
quarks. It is fully instrumented in the \mbox{pseudorapidity} range $2<\eta <5$ and has tracking capability in the backward direction, in the range $-3.5<\eta <-1.5$. The detector includes a high-precision tracking system
consisting of a silicon-strip vertex detector (VELO) surrounding the $pp$
interaction region, a large-area silicon-strip detector located
upstream of a dipole magnet with a bending power of about
$4{\rm\,Tm}$, and three stations of silicon-strip detectors and straw
drift tubes placed downstream of the magnet.
The tracking system provides a measurement of momentum, \ptot, of charged particles with
a relative uncertainty that varies from 0.5\% at low momentum to 1.0\% at 200\gevc.
Photons, electrons and hadrons are identified by a calorimeter system consisting of
scintillating-pad (SPD) and preshower detectors, an electromagnetic
calorimeter and a hadronic calorimeter. Muons are identified by a
system composed of alternating layers of iron and multiwire
proportional chambers.

The trigger consists of a hardware stage, based on information from the calorimeter and muon
systems, followed by a software stage, which applies a full event
reconstruction. The hardware trigger requires events to contain at least one muon with a \pt greater than 200\mevc. Low-multiplicity events are selected by requiring that fewer than ten hits should be detected in the scintillating pad detector, positioned just upstream of the electromagnetic calorimeter.
  In the subsequent software trigger, both of the final-state muons are required to have 
  $\pt>400\mevc$. 

The exclusive production of $\Upsilon(nS)$ resonances is simulated using the {\sc SuperChiC} software package~\cite{HarlandLang:2010ep}, which provides the four-momentum of a single, transversely polarised $\Upsilon(nS)$ resonance in each event. The decay of the $\Upsilon(nS)$ candidate is described by \evtgen~\cite{Lange:2001uf}, in which final-state radiation is generated using \photos~\cite{Golonka:2005pn}. The interaction of the generated particles with the detector, and its response,
are implemented using the \geant toolkit~\cite{Allison:2006ve, *Agostinelli:2002hh} as described in Ref.~\cite{LHCb-PROC-2011-006}. Samples, each containing one million events, are prepared for $\Upsilon(nS)$ resonances decaying to $\mup\mun$. In the same way, nine background samples of a similar size are prepared containing events where an exclusively produced $\chi_{b0,1,2}(1P,2P,3P)$ meson decays to the $\Upsilon(nS)\gamma$ final states with $\Upsilon\to\mup\mun$. Separate samples are prepared for every  $\chi_b(mP)\to\Upsilon(nS)\gamma$ $(m,n=1,2,3; n\leq m)$ decay.
\section{Candidate selection}
\label{sec:selectionAndEfficiency}
Selection criteria are applied offline to events that pass the trigger requirements, to select well-reconstructed $\Upsilon(nS)$ candidates and to ensure the absence of unrelated detector activity. The latter set of requirements favours events containing a single $pp$ interaction per bunch crossing.

The final-state tracks, which must be associated with hits in the muon chambers, are required to lie in the pseudorapidity range $2<\eta(\mu^\pm)<4.5$ and to be of good quality. In extracting the differential cross-section for the $\Upsilon(1S)$, the following intervals in $\Upsilon(1S)$ rapidity are considered: $2< y<3$, $3< y<3.5$ and $3.5< y<4.5$. Dimuon candidates are selected if the invariant mass falls in the range between 9\gevcc and 20\gevcc, and the candidate $\ptsq$ is less than 2$\gev^2/c^2$. The latter requirement favours photoproduction candidates, which have a characteristically low-$\pt$. Events are rejected if one or more tracks are reconstructed in the backward direction. In the forward region exactly two tracks, corresponding to the muon candidates, are required, and these must be reconstructed both in the VELO and in the downstream tracking detectors. 

The selection criteria affect not only the $\Upsilon(nS)$ candidate but also the level of activity in the rest of the event, specifically through the requirements that there should be exactly two forward tracks, no backward tracks and fewer than ten SPD hits. The event is excluded if more than one proton-proton interaction occurs, causing a larger number of additional SPD hits or extra tracks to be reconstructed. The probability for an exclusive $\Upsilon$ event not to feature additional activity from another $pp$ interaction in the same beam crossing is determined as the fraction of events containing no activity, according to these criteria, in a randomly accepted, hence unbiased, sample. After subtracting the contribution of the dimuon candidate, an event may contain fewer than eight SPD hits and no reconstructed tracks in the backward direction or tracks in the forward region. The fraction of randomly triggered events passing these criteria, $f_{\rm SI}$, is found to be  $(23.63 \pm 0.04)\%$ for the 7\tev data and $(18.48\pm 0.02)\%$ for the 8\tev data. The difference arises because of the different beam conditions in the two data-taking periods, leading to a different average number of proton-proton interactions per event.

The reconstruction, trigger and offline selection efficiencies are determined using simulated samples, and the combined efficiency varies between 77\% and 84\%.
For signal candidates that pass the trigger and reconstruction stages, the offline selection criteria are more than 99\% efficient.
\section{Determining the exclusive  yield}
\label{sec:Fitting}
Candidates reconstructed in the 7\tev and 8\tev data sets are combined in a single sample, and two unbinned, extended, maximum-likelihood fits are carried out. A first fit is performed to the dimuon invariant mass spectrum, between 9\gevcc and 20\gevcc. The fit contains a non-resonant background component and three resonant components. The three resonant components each receive contributions from exclusive signal, inelastic background and $\chi_b\to\Upsilon\gamma$ feed-down decays. These contributions are indistinguishable in the invariant mass distribution. 

The probability density function (PDF) used to model each $\Upsilon(nS)$ signal peak is a Gaussian function with modified tails (a double-sided crystal ball function~\cite{CB}). The mass differences for the $\Upsilon(2S){\rm-}\Upsilon(1S)$ and $\Upsilon(3S){\rm-}\Upsilon(1S)$ resonances are taken from Ref.~\cite{PDG2014}. The ratios of the $\Upsilon(2S)$ and $\Upsilon(3S)$ resolutions with respect to the $\Upsilon(1S)$ are fixed to the ratio of their masses with respect to the mass of the $\Upsilon(1S)$, following the procedure used in previous $\Upsilon$ measurements using LHCb data \cite{UpsilonLHCb7TeV}.
The parameters that govern the shapes of the tails are taken from simulation, as is the resolution of the $\Upsilon(1S)$ resonance, which varies from 35\mevcc to 57\mevcc in the different rapidity ranges. The yields of the signal components are all free to vary independently.

A background PDF accounting for the non-resonant background is modelled using an exponential shape where the slope and normalisation are allowed to vary.

\begin{figure}[tbp]
\centering
\includegraphics[width=.6\textwidth]{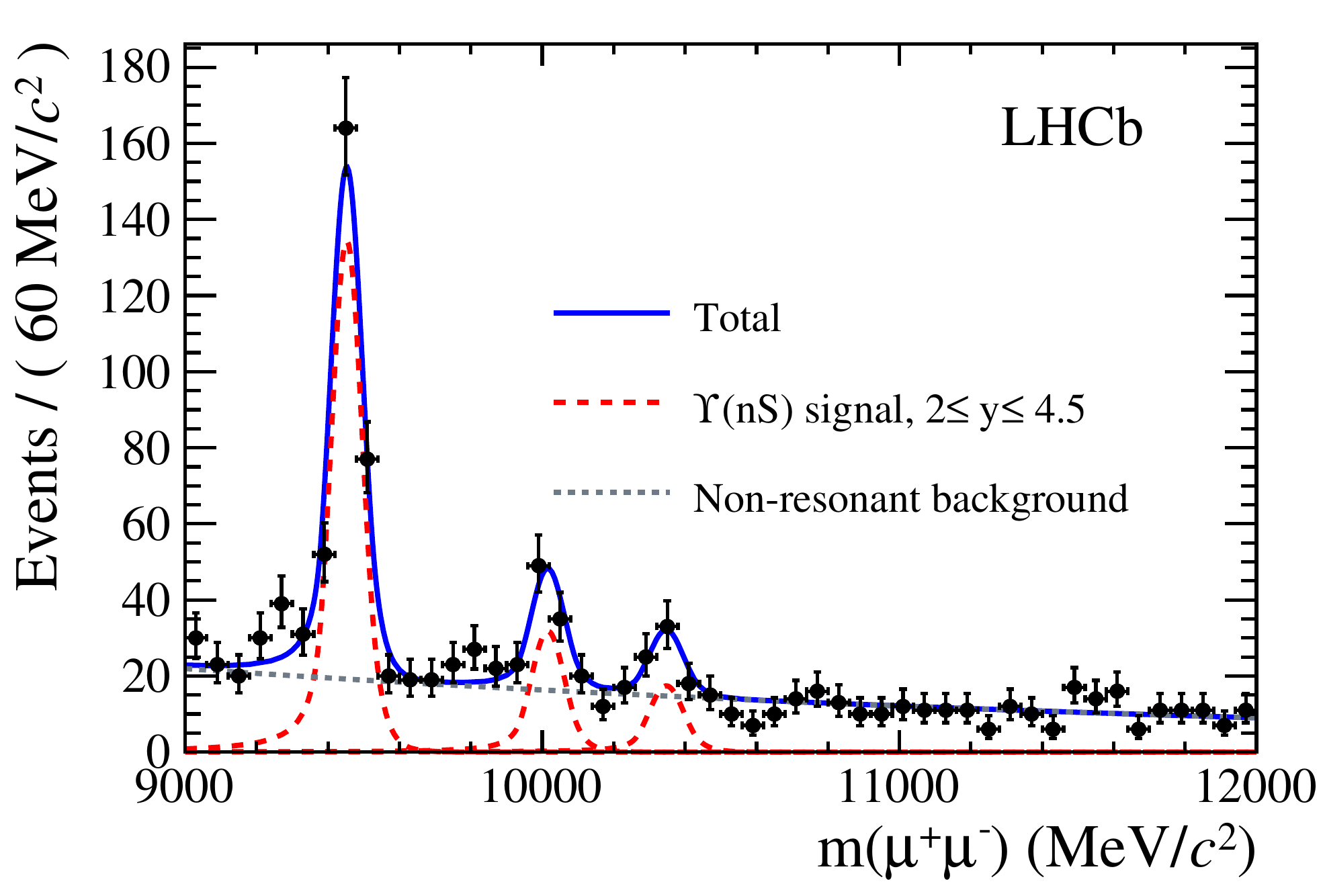}
\caption{Invariant dimuon mass spectrum for 7\,TeV and 8\,TeV data in the rapidity range $2< y(\Upsilon)<4.5$ (black points). The fit PDF is superimposed (solid blue line). The $\Upsilon(1S,2S,3S)$ signal components, used to derive weights, are indicated with a long-dashed (red) line, and the non-resonant background is marked with a short-dashed (grey) line.  \label{fig:InvMassFit}}
\end{figure}

The data are fitted in the whole rapidity range and in bins of rapidity. The fit results are given in Table~\ref{tab:InvMassFitFitPars} and the fit in the full rapidity range is shown between $9\gevcc$ and $12\gevcc$ in Fig.~\ref{fig:InvMassFit}. 

\begin{table}[tbp]
\centering
\caption{Results of the invariant mass fits, within each rapidity interval. \label{tab:InvMassFitFitPars}}
\begin{tabular}{l r@{ $\pm$ }l  r@{ $\pm$ }l r@{ $\pm$ }l r@{ $\pm$ }l}
\hline
Parameter & \multicolumn{2}{c}{$2< y<4.5$} & \multicolumn{2}{c}{$2< y<3$} & \multicolumn{2}{c}{$3< y<3.5$} & \multicolumn{2}{c}{$3.5< y<4.5$}\\
\hline
%%%% 
$\Upsilon(1S,2S,3S)$ yield & 382 & 26 & 146 & 16 & 133 & 16 & 94 & 14\\
\hspace{0.3cm}$\Upsilon(1S)$ fraction & 0.71 & 0.03 & 0.74 & 0.05 & 0.72 & 0.06 & 0.68 & 0.07\\
\hspace{0.3cm}$\Upsilon(2S)$ fraction & 0.18 & 0.03 & 0.16 & 0.04 & 0.15 & 0.05 & 0.26 & 0.06\\
$\Upsilon(1S)$ mass (\mevcc) & 9452.5 & 3.3 & 9453.2 & 4.3 & 9452.4 & 5.6 & 9452.0 & 9.0\\
%%%%%
\hline
\end{tabular}
\end{table}

Two sources of background contribute to the fitted signal: feed-down from $\chi_b\to\Upsilon\gamma$ decays, and inelastic interactions that involve the undetected products of proton dissociation or additional gluon radiation. 

The feed-down background is estimated using a combination of data and simulation, considering $\chi_b(mP)\to\Upsilon(nS)\gamma$ decays. Events are considered in the data set if exactly one photon is found in addition to the $\Upsilon$ candidate. Regions in the $\Upsilon\gamma$ invariant mass spectrum are defined, corresponding to the $\chi_b(1P,2P,3P)$ states, and the number of $\chi_b$ candidates, $N_{\chi_b}$, for each decay $\chi_b(mP)\to\Upsilon(nS)\gamma$ is counted. An estimate of the total feed-down content of the $\Upsilon$ data sample from each $\chi_b$ state is found using the expression:
\begin{equation}
N_{\textrm{feed-down, }\chi_b(mP)\to\Upsilon(nS)\gamma} = \frac{ N_{\chi_b} \times \mathcal{F} }{\epsilon_\gamma\times \epsilon_\textrm{mass-range}}.
\end{equation}
Here $\mathcal{F}$ is the purity of the $\Upsilon(nS)$ in the corresponding mass window with respect to the non-resonant $\mup\mun\gamma$ background, determined by fitting the dimuon mass spectrum for events with exactly one reconstructed photon; $\epsilon_\gamma$ is the efficiency for reconstructing the photon produced in each $\chi_b(mP)$ decay, determined using simulated exclusive $\chi_b(mP)\to\Upsilon(nS)\gamma$ decays; and $\epsilon_\textrm{mass-range} = 0.9$ corrects for the fraction of signal $\Upsilon$ candidates which are expected to fall outside the mass window. There are too few $\Upsilon(3S)\gamma$ candidates to estimate the purity precisely so it is assumed to be 100\%. Because of limited mass resolution and small sample sizes the $\chi_b$ spin states cannot be resolved, so equal contributions from the $\chi_{b1}(mP)$ and $\chi_{b2}(mP)$ states are assumed. The $\chi_{b0}$ radiative decay rate is expected to be relatively suppressed and is therefore neglected \cite{PDG2014}. The feed-down background yields are given in Table \ref{tab:ChibFeedDown}.

\begin{table}[tbp]
\centering\caption{Estimated yields of feed-down background from $\chi_b(mP)\to\Upsilon(nS)\gamma$ decays in each $\Upsilon(nS)$ sample, where the uncertainties are statistical only. \label{tab:ChibFeedDown}}
\begin{tabular}{l l  r@{ $\pm$ }l r@{ $\pm$ }l r@{ $\pm$ }l}
\hline
Signal window & $\Upsilon$ sample & \multicolumn{6}{l}{Estimated contamination yield}\\
			&				& \multicolumn{2}{l}{$\chi_b(1P)$} & \multicolumn{2}{l}{$\chi_b(2P)$} & \multicolumn{2}{l}{$\chi_b(3P)$} \\			
\hline
%%%% 
$2< y(\Upsilon)<4.5$
& $\Upsilon(1S)$ & 63 & 10 & 14 & 5 & 3 & 2\\
& $\Upsilon(2S)$ & \multicolumn{2}{c}{$-$}  & 43 & 12 & 5 & 3\\
& $\Upsilon(3S)$ & \multicolumn{2}{c}{$-$}  & \multicolumn{2}{c}{$-$}  & 21 & 21\\
\hline
$2< y(\Upsilon)<3$
& $\Upsilon(1S)$ & 31 & 8 & 2 & 2 & 0 & 2\\
$3< y(\Upsilon)<3.5$
& $\Upsilon(1S)$ & 22 & 6 & 10 & 4 & 0 & 2\\
$3.5< y(\Upsilon)<4.5$
& $\Upsilon(1S)$ & 8 & 4 & 0 & 2 & 3 & 2\\
%%%%%%
\hline
\end{tabular}
\end{table}

\begin{figure}[tbp]
\centering
\includegraphics[width=.6\textwidth]{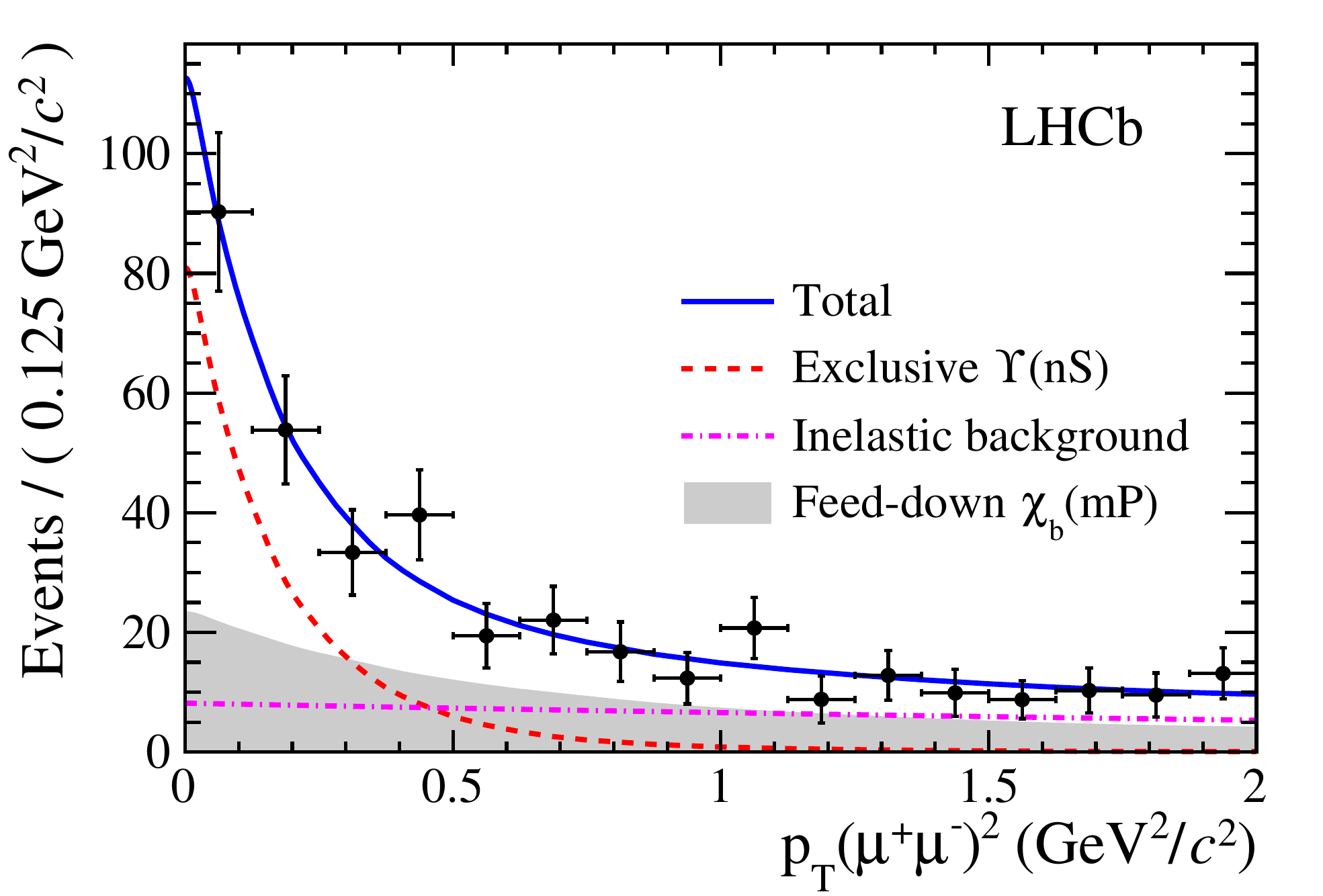}
\caption{Fit to the $\ptsq$ distribution of the $\Upsilon$ candidates in the full rapidity range. \label{fig:FullPurityFit}}
\end{figure}

Since the mass shapes for signal and background do not significantly depend on \pt over the \pt range considered, the $\ptsq$ distribution of the $\Upsilon$ candidates is determined using the $\sPlot$ technique \cite{sPlot}. A fit is then performed to the $\ptsq$ distribution, shown in Fig.~\ref{fig:FullPurityFit}, using candidates in the full rapidity range $2.0< y(\Upsilon)< 4.5$, with fit components corresponding to the $\Upsilon$ signal, inelastic background and feed-down background. The fraction of exclusive signal calculated from this fit is assumed to be the same for each rapidity bin.

The $\ptsq$ distribution for the exclusive signal is derived from the simulated sample. At HERA, the distribution of the exclusive charmonium signal as a function of the candidate $\ptsq$ was well described by an exponential function, $\exp(-b\ptsq)$ \cite{HERAJpsi2013,ZEUSJpsi2002}. The $\ptsq$ distribution provides discrimination among various production sources because it approximates the squared four-momentum transfer, $|t|$, which depends on the production mechanism. Following Ref.~\cite{RonanUpdatedJpsi}, Regge phenomenology is used to extrapolate the slope measured by HERA up to LHC energies according to the expression
\begin{equation}
b(W) = b_0+4\alpha'\log\left(\frac{W}{W_0}\right),
\end{equation}
where $\alpha'$ describes the slope of the exchange Regge trajectory and the constant $b_0$ is specific for interactions at a given photon-proton centre-of-mass energy, $W_0$ \cite{Jones:2013pga}. The {\sc SuperChiC} generator \cite{HarlandLang:2010ep} models the pomeron-photon exchange and performs this extrapolation. Since the only published measurement of $b_\Upsilon$ has very low precision \cite{ZEUSUpsilontDep}, the generator is tuned to reproduce the LHCb measurement of $b_{J/\psi} = 5.7\pm0.1$\,$\gev^{-2}c^2$ \cite{RonanUpdatedJpsi} in exclusive $J/\psi$ production. The input values to {\sc SuperChiC} are $b_0 = 5.6$\,$\gev^{-2}c^2$, $\alpha'=0.2$\,$\gev^{-2}c^2$ and $W_0=90$\gev. The fit PDF is obtained from the simulated samples with these inputs, and a kernel estimation is employed to derive a shape to fit to data \cite{KernelEstimation}. It is assumed that the inelastic background component is distributed according to a single exponential function \cite{RonanUpdatedJpsi}. The slope and yield of this background function are free to vary in the fit.

The total contamination from $\chi_b(mP)\to\Upsilon(nS)\gamma$ decays is constrained to be the sum of the contributions in Table~\ref{tab:ChibFeedDown} and enters the $\ptsq$ fit by means of a Gaussian constraint. Given that no analysis of exclusive $\chi_b$ production has been undertaken, and the consequent lack of knowledge of the exclusive purity of the very small sample of reconstructed $\chi_b$ candidates, it is assumed that inelastic processes contribute half of this feed-down background and the same inelastic background PDF is employed as that used to model inelastic $\Upsilon(nS)$ production. For the exclusive component, the shape of the dimuon $\ptsq$ PDF depends on the $\chi_b(mP)$ meson source, and the PDFs for each source, determined from simulation, are combined according to their relative contributions to the total feed-down yield. 

The exclusive purity, ${\cal P}$, is defined as the ratio of the exclusive signal yield, $N_{\rm exclusive}$, to the number of candidates remaining in the sample, {$N_{\rm exclusive}+N_{\rm inelastic}$}, after subtraction of the feed-down yield. The fit to the $\ptsq$ distribution in the full rapidity range, shown in Fig.~\ref{fig:FullPurityFit}, gives
\begin{equation}
{\cal P}\equiv \frac{N_{\rm exclusive}}{N_{\rm exclusive}+N_{\rm inelastic}} = (54\pm11)\,\%,\nonumber
\end{equation}
with the exponential slope of the inelastic background measured to be $-0.21\pm0.26 \gev^{-2}c^2$.  The results of fits to the $\ptsq$ distribution in each rapidity interval  are  consistent with this value. In order to validate the fit procedure, a set of pseudoexperiments is generated using the parameters obtained from the fit to the data, and the same fit is applied to each pseudoexperiment. The uncertainty on the purity is underestimated by 15\% in the fit and the statistical uncertainty quoted takes account of this.

\section{Systematic uncertainties}
\label{sec:Systematics}
The relative systematic uncertainties for the $\Upsilon(1S,2S,3S)$ cross-sections in the various rapidity ranges are summarised in Table~\ref{tab:SystematicsSummary}.

Contributions to the systematic uncertainty arising from the $\ptsq$ fit are considered: the uncertainty in the signal $\ptsq$ distribution as modelled by the {\sc SuperChiC} generator and the variation of the exclusive signal PDF expected in the various rapidity bins. The {\sc SuperChiC} generator is tuned to reproduce measurements of exclusive $J/\psi$ meson production made by LHCb~\cite{RonanUpdatedJpsi}. As no sufficiently precise measurements of the $\ptsq$ distribution in exclusive $\Upsilon(nS)$ resonance production exist, an estimate is made following Ref.~\cite{Jones:2013pga}, where it is argued from Regge theory that the slope $b_0$ of the proton  should be reduced by $4\alpha'\log (m_{\Upsilon(nS)} / m_{J/\psi})$. A simulated sample is generated accordingly and used to derive a signal $\ptsq$ template. Changing $b_0$ from 5.6 to 4.7 produces a relative decrease in the exclusive yields of 6\%, and this change is taken as the systematic uncertainty. For the differential cross-section measurements, the dependence of the signal $\ptsq$ shape on rapidity is studied by replacing the exclusive signal $\ptsq$ PDF with those determined in the smaller rapidity ranges, and the largest change in purity is taken as the uncertainty.  Combining the systematic uncertainties in quadrature yields a total uncertainty for the exclusive purity, ${\cal P}$, between 7.2\% and 8.2\%. In addition, the possibility for variation in the shape of the continuum dimuon background in $\ptsq$ as a function of mass is considered. The determination of the exclusive purity, $\mathcal{P}$, is repeated in the dimuon invariant mass range from 9 to $12\gevcc$, and the difference is taken as a conservative estimate of the systematic uncertainty.
 In Table~\ref{tab:SystematicsSummary} these sources contribute to the uncertainty denoted `purity fit'.
 
The uncertainty arising from the $\ptsq$ shape derived from simulation and used to describe the feed-down background is considered separately. The feed-down background PDF is constructed using only contributions from the $\chi_{b1}(mP)$ and $\chi_{b2}(mP)$ background components, in equal parts, assuming no contribution from $\chi_{b0}(mP)$ decays. Since it is not possible to resolve the spin states in data, we consider a conservative change where the nominal PDF is replaced with that for background originating from the decay of a $\chi_{b0}(mP)$ meson. The fit to data is repeated and the change in exclusive purity is taken as the associated uncertainty. In addition, there is uncertainty associated with the exclusive fraction of the $\chi_b(mP)$ feed-down background, which in turn affects the overall shape of the feed-down background used in the $\ptsq$ spectrum. Since the size of the data set is too small to allow a data-driven estimate of the $\chi_b(mP)$ sample exclusive purity, the PDF for this purity is assumed to be uniform between 0\% and 100\%, and the effect of changing it by $\pm 1$ standard deviation ($0.50\pm0.29$) is therefore considered. The resulting change in the exclusive $\Upsilon(nS)$ sample purity is taken as the systematic uncertainty. The estimate of the yield of $\chi_b(mP)$ meson feed-down in data includes the determination of the photon reconstruction efficiency using simulated samples of $\chi_b(mP)\to\Upsilon(nS)\gamma$ decays. Samples of $B^\pm\to J\psi K^{*\pm}(\to \Kpm\piz)$ and $\Bpm\to J/\psi \Kpm$ decays are used to validate the agreement between photon reconstruction efficiencies in data and simulation. An uncertainty of 5\% is taken as the systematic uncertainty on the photon reconstruction efficiency to account for the small differences seen \cite{Govorkova:2015vqa}. The resulting uncertainty is very small for the $\Upsilon(1S)$ but larger for the $\Upsilon(2S,3S)$ samples where the relative contamination from the $\chi_b(mP)$ background is larger.
These three systematic uncertainties on the cross-section are combined in quadrature and are presented as the `feed-down b.g.' systematic uncertainty in Table \ref{tab:SystematicsSummary}.

An estimate of the contamination of the $\Upsilon(1S)$ and $\Upsilon(2S)$ samples from the decays $\Upsilon(2S)\to\Upsilon(1S)\piz\piz$, $\Upsilon(3S)\to\Upsilon(2S) \{\piz\piz, \gamma\gamma\}$, is made using the observed $\Upsilon(nS)$ candidate yields in data and the relevant $\Upsilon'\to\Upsilon X$ and $\Upsilon^{(')}\to\mup\mun$ branching fractions \cite{PDG2014}. The estimated contaminations are taken as systematic uncertainties for the $\Upsilon(1S)$ and $\Upsilon(2S)$ cross-sections.

Uncertainties in the PDFs used to fit the $\Upsilon(nS)$ candidate invariant mass spectrum are considered. Alternative $\Upsilon(nS)$ signal PDFs are produced, obtained using kernel estimation. The systematic uncertainties on the exclusive purity and each of the yields are assessed using pseudoexperiments generated with the nominal invariant mass PDFs and fitted with a model where the signal PDFs are replaced by those obtained using kernel estimation. The effect of replacing the exponential PDF used to model the non-resonant background in the invariant mass fit with a second-order polynomial function is found in the same way using pseudoexperiments.
Combining these two sources of uncertainty in quadrature leads to a relative uncertainty which is less than 4\% for all the cross-sections. This uncertainty is labelled `mass fit' in Table \ref{tab:SystematicsSummary}.

The LHCb integrated luminosity has been measured with a relative uncertainty of 1.7\% at 7\tev and 1.2\% at 8\tev \cite{Aaij:2014ida}. The integrated luminosity is multiplied by the estimated, selection-dependent, fraction of events that contain no interactions other than the one that produces the signal candidate, $f_{\rm SI}$. The determination of $f_{\rm SI}$ from data depends upon the subtraction of the $\Upsilon$ signal candidate's SPD hits. The spread in the signal candidate SPD hit multiplicity is estimated from data to be one hit, and the fraction $f_{\rm SI}$ is therefore recomputed with the signal subtraction increased from two to three and the change is taken to be the systematic uncertainty. To account for variations as a function of data-taking time, the variation of the estimated single-interaction fraction is evaluated in each uninterrupted period of data-taking during which conditions are constant, typically an hour long, instead of considering each year as a whole, and the change with respect to the nominal fraction is taken as the systematic uncertainty. Combining the uncertainties in quadrature yields an overall relative uncertainty for each year of 2.3\%. The systematic uncertainties on the luminosity for each year are assumed to be 100\% correlated.

The branching fractions, $\mathcal{B}$, for $\Upsilon(1S)$, $\Upsilon(2S)$ and $\Upsilon(3S)$ to decay to the dimuon final state are accounted for to determine the $\Upsilon(nS)$ production cross-section. These branching fractions are taken from Ref.~\cite{PDG2014} and, for the  $\Upsilon(1S)$, $\Upsilon(2S)$ and $\Upsilon(3S)$ states, carry relative uncertainties of 2\%, 9\% and 10\%, respectively. These are propagated to an uncertainty on the production cross-section.

\begin{table}[tbp]
\centering
\caption{Summary of the relative systematic uncertainties, in \%.\label{tab:SystematicsSummary}}
\begin{tabular}{l| c c c | c c c}
\hline
 & $2< y<3$ & $3< y<3.5$ & $3.5< y<4.5$ & \multicolumn{3}{c}{$2< y<4.5$}\\
 & $\Upsilon(1S)$ & $\Upsilon(1S)$ & $\Upsilon(1S)$ & $\Upsilon(1S)$ & $\Upsilon(2S)$ & $\Upsilon(3S)$ \\
\hline
%%%
Purity fit 				& 14.2 					& 14.2 			& 14.2 			& 13.7 
& 13.7 				& 13.7\\
Feed-down b.g. 		& 12.2 					& 12.2 			& 12.3 			& 12.2 			
& 14.6 				& 12.5\\
$\Upsilon'$ feed-down 	& \phantom{0}4.0 			& \phantom{0}4.3 	& \phantom{0}5.4	& \phantom{0}4.5  
& 11.1 				& \phantom{0}$-$\\
Mass fit 				& \phantom{0}2.2 			& \phantom{0}2.8 	& \phantom{0}2.9	& \phantom{0}2.1   
& \phantom{0}2.8 		& \phantom{0}3.6\\
Luminosity			& \phantom{0}2.3 			& \phantom{0}2.3 	& \phantom{0}2.3	& \phantom{0}2.3  
& \phantom{0}2.3	   	& \phantom{0}2.3\\
$\mathcal{B}(\Upsilon\to\mup\mun)$ &  \phantom{0}2.0 	& \phantom{0}2.0 	& \phantom{0}2.0	& \phantom{0}2.0  
& \phantom{0}8.8	   	& \phantom{0}9.6\\
\hline
Total					&  19.5 					& 19.7 			& 20.0			& 19.3 
& 24.8	   	& 21.4\\
%%% 
\hline
\end{tabular}
\end{table}

\section{Cross-section}
\label{sec:CrossSection}
The cross-section is obtained using
\begin{equation}
\sigma = \frac{ N_{\rm exclusive} }{ {\cal L}\times \epsilon \times \mathcal{B}(\Upsilon(nS)\to\mu^+\mu^-)}.
\end{equation}
The effective integrated luminosity, $\cal{L}$, is 580\invpb, taking into account the values of $f_\textrm{SI}$ given in Sect.~\ref{sec:selectionAndEfficiency} \cite{Aaij:2014ida}. 

 The quantity $\epsilon$ is the efficiency correction, which is obtained in each rapidity bin and for each resonance, and which is averaged for 7\tev and 8\tev data according to the luminosity in each year, and $\mathcal{B}(\Upsilon(nS)\to\mu^+\mu^-)$ is the $\Upsilon(nS)\to\mup\mun$ branching fraction. The measured exclusive production cross-sections in the LHCb acceptance are
\begin{eqnarray}
\sigma(pp\to p\Upsilon(1S)p) &=& 9.0 \pm 2.1 \pm 1.7\textrm{ pb},\nonumber\\
\sigma(pp\to p\Upsilon(2S)p) &=& 1.3 \pm 0.8 \pm 0.3\textrm{ pb},\textrm{ and}\nonumber\\
\sigma(pp\to p\Upsilon(3S)p) &<& 3.4\textrm{ pb} \textrm{ at the 95\% confidence level},\nonumber
\end{eqnarray}
where the first uncertainties are statistical and the second are systematic, and where the limit on $\sigma(\Upsilon(3S))$ includes possible contamination from $\chi_b$ feed-down. The limit is calculated using pseudo-experiments and includes the effect of systematic uncertainties, where correlations are assumed to be negligible. The $\Upsilon(1S)$ production cross-section is given in smaller ranges of $\Upsilon(1S)$ rapidity in Table~\ref{tab:FinalCrossSection}. 

\begin{table}[tbp]
\centering
\caption{Production cross-section for the $\Upsilon(1S)$ resonance in ranges of $\Upsilon(1S)$ rapidity, where the muons are required to lie in the pseudorapidity range $2<\eta(\mu^\pm)<4.5$. The first uncertainties are statistical and the second are systematic.\label{tab:FinalCrossSection}}
\begin{tabular}{l  c c c}
\hline
 & $2< y<3$ & $3< y<3.5$ & $3.5< y<4.5$\\
%%%% 
$\sigma(\Upsilon(1S))$ (pb)  & 3.4 $\pm$ 0.9 $\pm$ 0.7 & 2.9 $\pm$ 0.8 $\pm$ 0.6 & 2.6 $\pm$ 0.8 $\pm$ 0.5\\
%%%%
\hline
\end{tabular}
\end{table}

\begin{figure}[tbp]
\centering
\includegraphics[width=.49\textwidth]{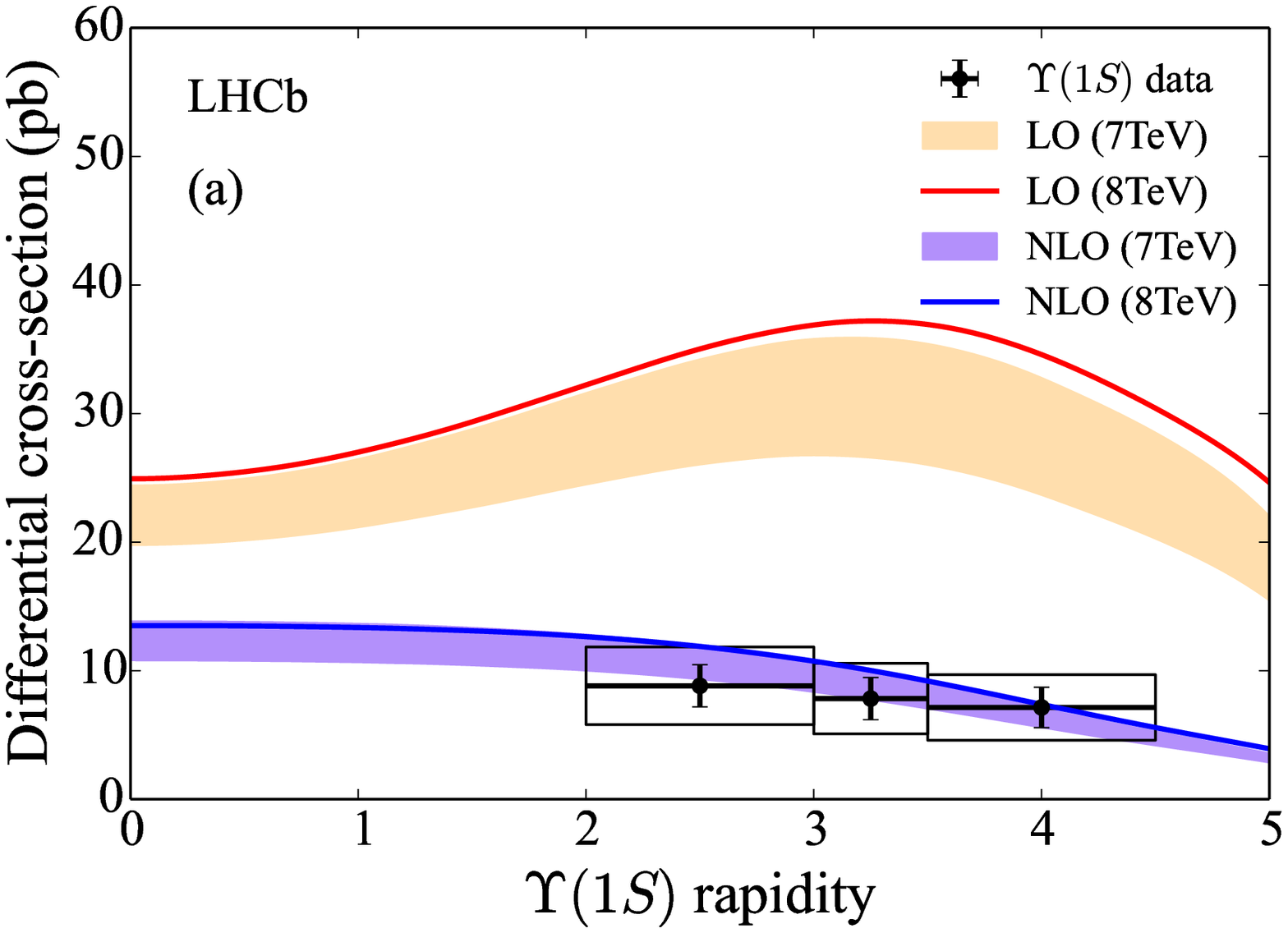}
\includegraphics[width=.5\textwidth]{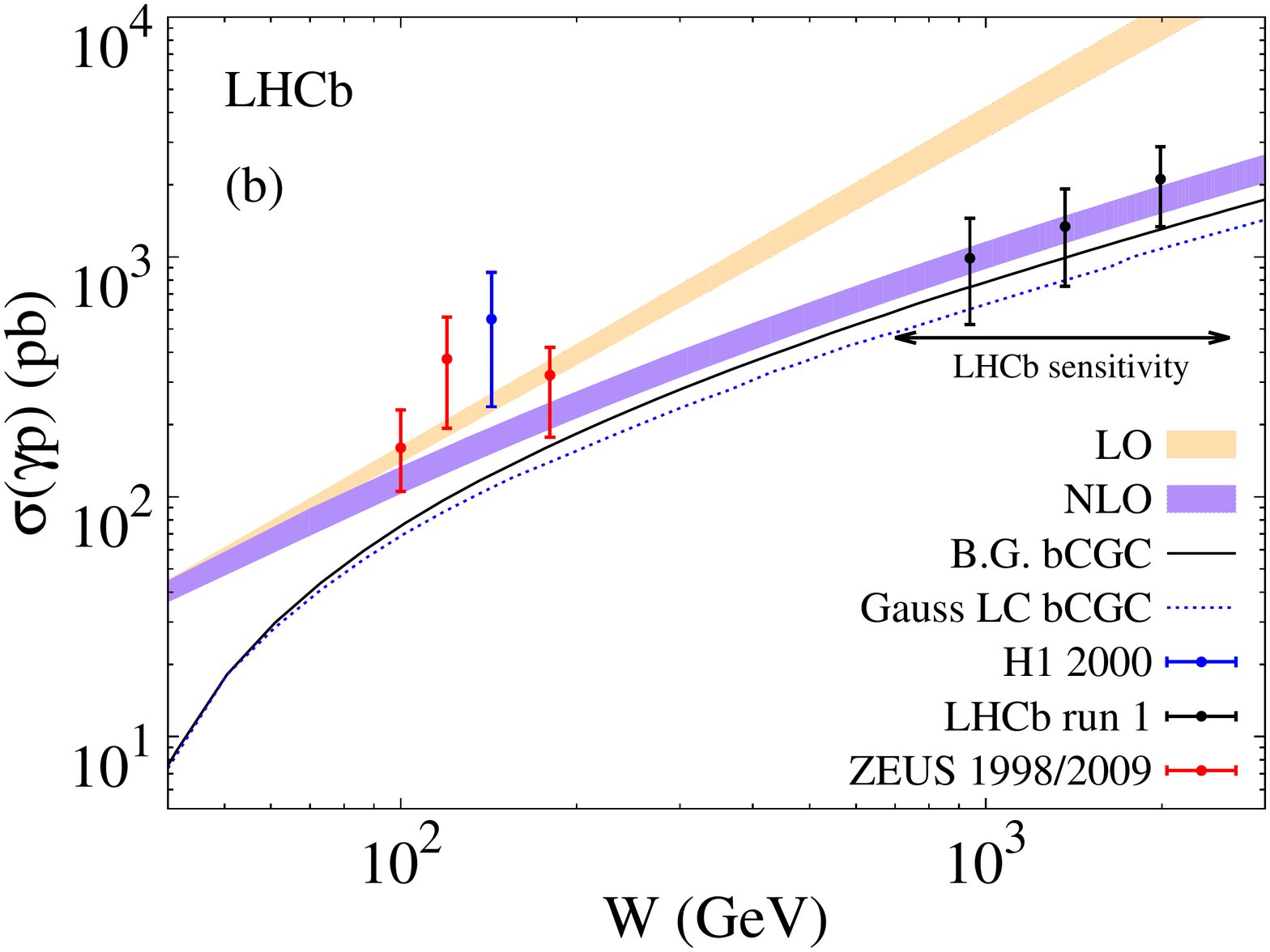}
\caption{Measurements of exclusive $\Upsilon(1S)$ photoproduction compared to theoretical predictions. In (a), the $\Upsilon(1S)$ cross-section in bins of rapidity is shown, compared to LO and NLO predictions. The LHCb measurements are indicated by black points with error bars for uncorrelated errors, and solid rectangles indicating the total uncertainty. In (b), the photon-proton cross-sections extracted from the LHCb results are indicated by black points, where the statistical and systematic uncertainties are combined in quadrature. The entire $W$-region in which these LHCb measurements are sensitive is indicated. Measurements made by H1 and ZEUS in the low-$W$ region are indicated by red and blue markers, respectively \cite{ZEUS1998,HERAJpsiAndUpsilonExclProduction2000,ZEUS2009}. Predictions from Ref.~\cite{Jones:2013pga} are included, resulting from LO and NLO fits to exclusive $J/\psi$ production data. The filled bands indicate the theoretical uncertainties on the 7 TeV prediction and the solid lines indicate the central values of the predictions for 8 TeV. In (b) predictions from Ref.~\cite{GoncalvesUpsilon2014} using different models for the $\Upsilon(1S)$ wave function are included, indicated by `bCGC'.\label{im:Prediction}}
\end{figure}

After correction for the LHCb geometrical acceptance, the cross-sections in Table~\ref{tab:FinalCrossSection} can be compared to theoretical predictions. The efficiency for an $\Upsilon$ candidate to be produced in the range $2< y(\Upsilon)<4.5$, and to decay to muons which lie inside the acceptance, $2<\eta(\mu^\pm)<4.5$, is 45\%. In the smaller ranges of $\Upsilon(1S)$ rapidity considered for the differential cross-section measurement, the efficiency is lowest in the outer ranges, at 39\% ($2<y<3$) and 36\% ($3.5<y<4.5$), and highest in the central range, at around 74\% ($3<y<3.5$). The correction depends on the rapidity distribution in the simulated sample, which has a different shape to those of, for example, the predictions in Fig.~\ref{im:Prediction}a. To estimate the systematic uncertainty on the geometrical acceptance correction, the simulated samples are reweighted to obtain a uniform rapidity distribution within each rapidity bin, and the change in the geometrical acceptance is taken as the systematic uncertainty. This corresponds to a relative change in the geometrical acceptance of less than 6\%. The differential cross-sections, $d\sigma(\Upsilon(1S))/dy$ are given in Table~\ref{tab:FinalCrossSectionScaled} and are shown in {Fig.~\ref{im:Prediction}a} compared to LO and NLO predictions~\cite{Jones:2013pga}. The LHCb data are in good agreement with the NLO prediction.

\begin{table}[tbp]
\centering
\caption{Measured $d\sigma(\Upsilon(1S))/dy$, where the data have been corrected for the effect of the LHCb geometrical acceptance.  The statistical and systematic uncertainties are combined in quadrature.\label{tab:FinalCrossSectionScaled}}
\begin{tabular}{l c c c}
\hline
 & $2< y<3$ & $3< y<3.5$ & $3.5< y<4.5$\\
%%%% 
$d\sigma(\Upsilon(1S))/dy$ (pb)  & 8.8 $\pm$ 3.0 & 7.8 $\pm$ 2.7 & 7.1 $\pm$ 2.6\\
%%%%
\hline
\end{tabular}
\end{table}

The LHCb results are also compared to theoretical predictions for the underlying photon-proton cross-section, as a function of the centre-of-mass energy of the photon-proton system, $W$, as shown in Fig.~\ref{im:Prediction}b. There are two contributions to the photoproduction of an $\Upsilon(nS)$ resonance, depending on which proton emits the virtual photon. The $pp$ cross-section is given by
\begin{equation}
\frac{d\sigma^{\textrm{th}}(pp\to p\Upsilon(1S)p)}{dy}   =    S^2(W_+)\left( k_+ \frac{dn}{dk_+} \right) \sigma_+^{\textrm{th}}(\gamma p) + S^2(W_-)\left( k_- \frac{dn}{dk_-} \right) \sigma_-^{\textrm{th}}(\gamma p),
\end{equation}
where the predictions for the photon-proton cross-section are weighted by absorptive corrections $S^2(W_\pm)$ and the photon fluxes $\frac{dn}{dk_\pm}$ for photons of energy $k_\pm  \approx (M_{\Upsilon(nS)}/2)\exp(\pm|y|)$. The absorptive corrections and photon fluxes are computed following Ref.~\cite{Jones:2013pga}.

The three bins of $\Upsilon(1S)$ rapidity chosen in this analysis correspond to ranges of $W$ for the $W_+$ and $W_-$ solutions. The contribution to the total cross-section from the $W_-$ solutions is expected to be small and is therefore neglected. The dominant $W_+$ solutions are therefore estimated assuming that they dominate the cross-section, and are shown in Fig.~\ref{im:Prediction}b. The magnitude of the theoretical prediction for the $W_-$ solutions is added as a systematic uncertainty. The good agreement with the NLO prediction seen in Fig.~\ref{im:Prediction}a is reproduced. The LHCb measurements probe a new kinematic region complementary to that studied at HERA \cite{ZEUS1998,HERAJpsiAndUpsilonExclProduction2000,ZEUS2009}, as seen in Fig.~\ref{im:Prediction}b, and discriminate between LO and NLO predictions. In Fig.~\ref{im:Prediction}b, the LHCb data are also compared to the  predictions given in Ref.~\cite{GoncalvesUpsilon2014} using models conforming to the colour glass condensate (CGC) formalism \cite{JalilianMarian:1996xnCGC} that take into account the $t$-dependence of the differential cross-section. All agree well with the data. The solid (black) and dotted (blue) lines correspond to two different models for the scalar part of the vector-meson wave function.

\section{Conclusion}
The first measurement of exclusive $\Upsilon(nS)$ production in $pp$ collisions at 7 and 8\tev is presented, and a differential cross-section is extracted as a function of $\Upsilon(1S)$ candidate rapidity. The data probe a previously unexplored kinematic region in photon-proton centre-of-mass energy. The results are compared to theoretical predictions and a strong preference for those including next-to-leading order calculations is seen. 
Exclusive production studies at LHCb will be improved during LHC Run II following the installation of scintillators at high $|\eta|$, which will allow for improved trigger efficiency for exclusive production processes and additional suppression of the background from inelastic interactions \cite{Herschel}.

% Do not include this in analysis note and conference reports
\section*{Acknowledgements}
\noindent The authors gratefully acknowledge discussions with Lucian Harland-Lang concerning the implementation and configuration of {\sc SuperChiC} to model the exclusive production process. We express our gratitude to our colleagues in the CERN
accelerator departments for the excellent performance of the LHC. We
thank the technical and administrative staff at the LHCb
institutes. We acknowledge support from CERN and from the national
agencies: CAPES, CNPq, FAPERJ and FINEP (Brazil); NSFC (China);
CNRS/IN2P3 (France); BMBF, DFG, HGF and MPG (Germany); INFN (Italy); 
FOM and NWO (The Netherlands); MNiSW and NCN (Poland); MEN/IFA (Romania); 
MinES and FANO (Russia); MinECo (Spain); SNSF and SER (Switzerland); 
NASU (Ukraine); STFC (United Kingdom); NSF (USA).
The Tier1 computing centres are supported by IN2P3 (France), KIT and BMBF 
(Germany), INFN (Italy), NWO and SURF (The Netherlands), PIC (Spain), GridPP 
(United Kingdom).
We are indebted to the communities behind the multiple open 
source software packages on which we depend. We are also thankful for the 
computing resources and the access to software R\&D tools provided by Yandex LLC (Russia).
Individual groups or members have received support from 
EPLANET, Marie Sk\l{}odowska-Curie Actions and ERC (European Union), 
Conseil g\'{e}n\'{e}ral de Haute-Savoie, Labex ENIGMASS and OCEVU, 
R\'{e}gion Auvergne (France), RFBR (Russia), XuntaGal and GENCAT (Spain), Royal Society and Royal
Commission for the Exhibition of 1851 (United Kingdom).

%\input{appendix}

% This should be taken out in the final paper
%\input{supplementary-app}

\addcontentsline{toc}{section}{References}
\setboolean{inbibliography}{true}
\bibliographystyle{LHCb}
\bibliography{main,LHCb-PAPER,LHCb-CONF,LHCb-DP,LHCb-TDR}

\newpage

% Author List ----------------------------                                                                                                                                                                                                                                                                                                
%  You need to get a new author list!                                                                                                                                                                                                                                                                                                    

\newpage
%%%%%%%%%%%%%%%%%%%%%%%%%%%%%%%%%%%%%%%%%%
\centerline{\large\bf LHCb collaboration}
\begin{flushleft}
\small
R.~Aaij$^{38}$, 
B.~Adeva$^{37}$, 
M.~Adinolfi$^{46}$, 
A.~Affolder$^{52}$, 
Z.~Ajaltouni$^{5}$, 
S.~Akar$^{6}$, 
J.~Albrecht$^{9}$, 
F.~Alessio$^{38}$, 
M.~Alexander$^{51}$, 
S.~Ali$^{41}$
% ...

{\footnotesize \it
% ...
$ ^{5}$Clermont Universit\'{e}, Universit\'{e} Blaise Pascal, CNRS/IN2P3, LPC, Clermont-Ferrand, France\\
$ ^{6}$CPPM, Aix-Marseille Universit\'{e}, CNRS/IN2P3, Marseille, France\\
% ...
$ ^{9}$Fakult\"{a}t Physik, Technische Universit\"{a}t Dortmund, Dortmund, Germany\\
% ...
$ ^{37}$Universidad de Santiago de Compostela, Santiago de Compostela, Spain\\
$ ^{38}$European Organization for Nuclear Research (CERN), Geneva, Switzerland\\
% ...
$ ^{41}$Nikhef National Institute for Subatomic Physics, Amsterdam, The Netherlands\\
% ...
$ ^{46}$H.H. Wills Physics Laboratory, University of Bristol, Bristol, United Kingdom\\
% ...
$ ^{51}$School of Physics and Astronomy, University of Glasgow, Glasgow, United Kingdom\\
$ ^{52}$Oliver Lodge Laboratory, University of Liverpool, Liverpool, United Kingdom\\
% ...
}
\end{flushleft}
%%%%%%%%%%%%%%%%%%%%%%%%%%%%%%%%%%%%%%%%%%

%\input{LHCb_authorlist.tex}

\end{document}